\documentclass[useAMS,usenatbib]{mn2e}
\usepackage{xspace}
\usepackage{graphicx}
\newcommand{\ignore}[1]{}
\providecommand{\ao}{}
\renewcommand{\ao}{adaptive optics (AO)\renewcommand{\ao}{AO\xspace}\renewcommand{\Ao}{AO\xspace}\xspace}
\newcommand{\Ao}{Adaptive optics (AO)\renewcommand{\ao}{AO\xspace}\renewcommand{\Ao}{AO\xspace}\xspace}
\newcommand{\wfs}{wavefront sensor (WFS)\renewcommand{\wfs}{WFS\xspace}\renewcommand{\wfss}{WFSs\xspace}\xspace}
\newcommand{\wfss}{wavefront sensors (WFSs)\renewcommand{\wfs}{WFS\xspace}\renewcommand{\wfss}{WFSs\xspace}\xspace}
\newcommand{\shwfs}{Shack-Hartmann \wfs (SHWFS)\renewcommand{\shwfs}{SHWFS\xspace}\xspace}
\newcommand{\dm}{deformable mirror (DM)\renewcommand{\dm}{DM\xspace}\renewcommand{\dms}{DMs\xspace}\renewcommand{\Dms}{DMs\xspace}\renewcommand{\Dm}{DM\xspace}\xspace}
\newcommand{\dms}{deformable mirrors (DMs)\renewcommand{\dm}{DM\xspace}\renewcommand{\dms}{DMs\xspace}\renewcommand{\Dms}{DMs\xspace}\renewcommand{\Dm}{DM\xspace}\xspace}
\newcommand{\Dms}{Deformable mirrors (DMs)\renewcommand{\dm}{DM\xspace}\renewcommand{\dms}{DMs\xspace}\renewcommand{\Dms}{DMs\xspace}\renewcommand{\Dm}{DM\xspace}\xspace}
\newcommand{\Dm}{Deformable mirror (DM)\renewcommand{\dm}{DM\xspace}\renewcommand{\dms}{DMs\xspace}\renewcommand{\Dms}{DMs\xspace}\renewcommand{\Dm}{DM\xspace}\xspace}

\newcommand{\shs}{Shack-Hartmann sensor (SHS)\renewcommand{\shs}{SHS\xspace}\renewcommand{\shss}{SHSs\xspace}\xspace}
\newcommand{\shss}{Shack-Hartmann sensors (SHSs)\renewcommand{\shs}{SHS\xspace}\renewcommand{\shss}{SHSs\xspace}\xspace}
\newcommand{\lgs}{laser guide star (LGS)\renewcommand{\lgs}{LGS\xspace}\renewcommand{\lgss}{LGSs\xspace}\xspace}
\newcommand{\lgss}{laser guide stars (LGSs)\renewcommand{\lgs}{LGS\xspace}\renewcommand{\lgss}{LGSs\xspace}\xspace}
\newcommand{\ngs}{natural guide star (NGS)\renewcommand{\ngs}{NGS\xspace}\renewcommand{\ngss}{NGSs\xspace}\xspace}
\newcommand{\ngss}{natural guide stars (NGSs)\renewcommand{\ngs}{NGS\xspace}\renewcommand{\ngss}{NGSs\xspace}\xspace}
\newcommand{\mems}{Micro-Electro-Mechanical Systems (MEMS)\renewcommand{\mems}{MEMS\xspace}\xspace}
\newcommand{\snr}{signal to noise ratio (SNR)\renewcommand{\snr}{SNR\xspace}\xspace}
\newcommand{\moao}{multi-object \ao (MOAO)\renewcommand{\moao}{MOAO\xspace}\xspace}
\newcommand{\mcao}{multi-conjugate adaptive optics (MCAO)\renewcommand{\mcao}{MCAO\xspace}\xspace}
\newcommand{\ltao}{laser tomographic adaptive optics (LTAO)\renewcommand{\ltao}{LTAO\xspace}\xspace}
\newcommand{\cpu}{central processing unit (CPU)\renewcommand{\cpu}{CPU\xspace}\renewcommand{\cpus}{CPUs\xspace}\xspace}
\newcommand{\cpus}{central processing units (CPUs)\renewcommand{\cpu}{CPU\xspace}\renewcommand{\cpus}{CPUs\xspace}\xspace}
\newcommand{\psf}{point spread function (PSF)\renewcommand{\psf}{PSF\xspace}\renewcommand{\psfs}{PSFs\xspace}\xspace}
\newcommand{\psfs}{point spread functions (PSFs)\renewcommand{\psf}{PSF\xspace}\renewcommand{\psfs}{PSFs\xspace}\xspace}
\newcommand{\fpga}{field programmable gate array (FPGA)\renewcommand{\fpga}{FPGA\xspace}\renewcommand{\fpgas}{FPGAs\xspace}\xspace}
\newcommand{\fpgas}{field programmable gate arrays (FPGAs)\renewcommand{\fpga}{FPGA\xspace}\renewcommand{\fpgas}{FPGAs\xspace}\xspace}
\newcommand{\sor}{successive over-relaxation (SOR)\renewcommand{\sor}{SOR\xspace}\xspace}
\newcommand{\fdpcg}{Fourier domain pre-conditioned gradient (FDPCG)\renewcommand{\fdpcg}{FDPCG\xspace}\xspace}
\newcommand{\map}{maximum a-posteriori (MAP)\renewcommand{\map}{MAP\xspace}\xspace}
\newcommand{\elt}{Extremely Large Telescope (ELT)\renewcommand{\elt}{ELT\xspace}\renewcommand{\elts}{ELTs\xspace}\xspace}
\newcommand{\elts}{Extremely Large Telescopes (ELTs)\renewcommand{\elt}{ELT\xspace}\renewcommand{\elts}{ELTs\xspace}\xspace}
\newcommand{\dugall}{Durham University generalised adaptive optics laser laboratory (DUGALL)\renewcommand{\dugall}{DUGALL\xspace}\xspace}
\newcommand{\fwhm}{full-width at half-maximum (FWHM)\renewcommand{\fwhm}{FWHM\xspace}\xspace}
\newcommand{\wht}{William Herschel Telescope (WHT)\renewcommand{\wht}{WHT\xspace}\xspace}
\newcommand{\emccd}{electron multiplying CCD (EMCCD)\renewcommand{\emccd}{EMCCD\xspace}\xspace}
\newcommand{\dasp}{Durham \ao simulation platform (DASP)\renewcommand{\dasp}{DASP\xspace}\xspace}
\newcommand{\eelt}{European \elt (E-ELT)\renewcommand{\eelt}{E-ELT\xspace}\xspace}
\newcommand{\mpi}{Message Passing Interface (MPI)\renewcommand{\mpi}{MPI\xspace}\xspace}
\newcommand{\smp}{symmetric multi-processing (SMP)\renewcommand{\smp}{SMP\xspace}\xspace}
\newcommand{\svd}{singular value decomposition (SVD)\renewcommand{\svd}{SVD\xspace}\xspace}
\newcommand{\gpu}{graphical processing unit (GPU)\renewcommand{\gpu}{GPU\xspace}\renewcommand{\gpus}{GPUs\xspace}\xspace}
\newcommand{\gpus}{graphical processing units (GPUs)\renewcommand{\gpu}{GPU\xspace}\renewcommand{\gpus}{GPUs\xspace}\xspace}
\newcommand{\fft}{fast Fourier transform (FFT)\renewcommand{\fft}{FFT\xspace}\xspace}
\newcommand{\ifu}{integral field unit (IFU)\renewcommand{\ifu}{IFU\xspace}\xspace}
\newcommand{\darc}{the Durham adaptive optics real-time controller (DARC)\renewcommand{\darc}{DARC\xspace}\renewcommand{\Darc}{DARC\xspace}\xspace}
\newcommand{\Darc}{The Durham adaptive optics real-time controller (DARC)\renewcommand{\darc}{DARC\xspace}\renewcommand{\Darc}{DARC\xspace}\xspace}
\newcommand{\cots}{commercial off-the-shelf (COTS)\renewcommand{\cots}{COTS\xspace}\xspace}
\newcommand{\rtcp}{real-time control pipeline (RTCP)\renewcommand{\rtcp}{RTCP\xspace}\xspace}
\newcommand{\rms}{root-mean-square (RMS)\renewcommand{\rms}{RMS\xspace}\xspace}
\newcommand{\sFPDP}{serial Front Panel Data Port (sFPDP)\renewcommand{\sFPDP}{sFPDP\xspace}\xspace}
\newcommand{\wpu}{wavefront processing unit (WPU)\renewcommand{\wpu}{WPU\xspace}\xspace}
\newcommand{\canary}{CANARY\xspace}

\newcommand{\rtcs}{real-time control system (RTCS)\renewcommand{\rtcs}{RTCS\xspace}\xspace}
\newcommand{\ptp}{point-to-point (PTP)\renewcommand{\ptp}{PTP\xspace}\xspace}
\newcommand{\sse}{streaming SIMD extension (SSE)\renewcommand{\sse}{SSE\xspace}\xspace}
\newcommand{\api}{application programming interface (API)\renewcommand{\api}{API\xspace}\xspace}
\newcommand{\corba}{Common Object Request Broker Architecture (CORBA)\renewcommand{\corba}{CORBA\xspace}\xspace}
\newcommand{\lqg}{linear quadratic gaussian (LQG)\renewcommand{\lqg}{LQG\xspace}\xspace}
\newcommand{\scao}{single conjugate adaptive optics (SCAO)\renewcommand{\scao}{SCAO\xspace}\xspace}
\newcommand{\dma}{direct memory access (DMA)\renewcommand{\dma}{DMA\xspace}\xspace}
\newcommand{\xao}{extreme adaptive optics (XAO)\renewcommand{\xao}{XAO\xspace}\xspace}
\newcommand{\vlt}{Very Large Telescope (VLT)\renewcommand{\vlt}{VLT\xspace}\xspace}
\newcommand{\sparta}{Standard Platform for Advanced Real-Time
  Applications (SPARTA)\renewcommand{\sparta}{SPARTA\xspace}\xspace}
\newcommand{\eso}{European Southern Observatory (ESO)\renewcommand{\eso}{ESO\xspace}\xspace}

\newcommand{\epics}{Exo-Planet Imaging Camera and Spectrograph (EPICS)\renewcommand{\epics}{EPICS\xspace}\xspace}
\newcommand{\iir}{infinite impulse response (IIR)\renewcommand{\iir}{IIR\xspace}\xspace}
\newcommand{\gtc}{Gran Telescopio Canarias (GTC)\renewcommand{\gtc}{GTC\xspace}\xspace}
\newcommand{\cog}{centre of gravity (CoG)\renewcommand{\cog}{CoG\xspace}\xspace}

\title[AO Correlation reference update]{Real-time correlation
  reference update for astronomical adaptive optics}
\author[A. G. Basden]{A. G. Basden$^{1}$\thanks{E-mail:
    a.g.basden@durham.ac.uk (AGB)}\\
$^{1}$Department of Physics, South Road, Durham, DH1 3LE, UK}

\begin{document}
\maketitle

\begin{abstract}
The use of laser guide stars in astronomical adaptive optics results
in elongated Shack-Hartmann wavefront sensor image patterns.  Image
correlation techniques can be used to determine local wavefront slope
by correlating each sub-aperture image with its expected elongated
shape, or reference image.  Here, we present a technique which allows
the correlation reference images to be updated while the adaptive
optics loop is closed.  We show that this can be done without
affecting the resulting point spread functions.  On-sky demonstration
is reported.  We compare different techniques for obtaining the
reference images, and investigate performance over a wide range of
adaptive optics system parameters.  We find that image correlation
techniques perform better than the standard centre-of-gravity
algorithm and are highly suited for use with open-loop multiple object
adaptive optics systems.
\end{abstract}
\begin{keywords}
Instrumentation: adaptive optics, techniques: image processing,
instrumentation: high angular resolution
\end{keywords}

\ignore{
Key question:  How does this differentiate between atmosphere and Na
layer variability?  Is it a case of time averaging?  Or of averaging
over sub-apertures?  Or can something be done on a NGS?  How far will
the technique proposed here get us?

It is useful because it allows us to maximise signal when spot shape
changes.  Which can happen regularly.  But, if we don't average for
long enough, or if the Na layer changes too regularly, then it will
produce wrong slope estimates.

So, how long is a good time to average for?

Maybe a plot of slope variance as a function of frames summed would be
good - it would tend to an asymtote, at which point this is the number
of frames required to average the atmosphere.

Need to say why O(1) second for averaging, and back this up.

Also introduce Rayleigh - how useful there.  eg extending range gate.

}

\section{Introduction}
All ground-based astronomical telescopes perform science by observing
through the Earth's atmosphere, which has a degrading effect on the
images obtained.  \Ao \citep{adaptiveoptics} is a technology employed
on most major telescopes, which seeks to remove some of the effects of
atmospheric turbulence, producing clearer, high resolution science
images as a result.  It is a crucial technology for the next
generation \elt facilities which will spend the significant majority
of their time producing \ao corrected observations.

The field of view of high image quality obtained from a classical \ao
system \citep{adaptiveoptics} (around the guide star that is used to
sense the atmospheric turbulence), is limited by the atmospheric
isoplanatic patch size, typically a few seconds of arc in diameter.
This results in degrading image quality for science targets away from
this guide star position.  Wide field adaptive optics systems have
recently been commissioned (for example, CANARY, \citet{canaryresultsshort})
which increase the corrected field of view by using multiple guide
stars and performing a tomographic reconstruction of the atmospheric
turbulence.  Herein lies a problem: there are only a limited number of
stars with sufficient brightness that form asterisms with the small
angular separation required for wavefront sensing for wide field \ao.
Therefore, sky coverage of such systems is limited.

Sky coverage is improved by the use of \lgss \citep{laserguidestar},
which allow artificial star asterisms to be placed anywhere on the sky
though sky coverage is still limited by the requirement that at least
one \ngs is necessary to overcome tip-tilt uncertainty.

\begin{figure}
\includegraphics[width=0.3\linewidth]{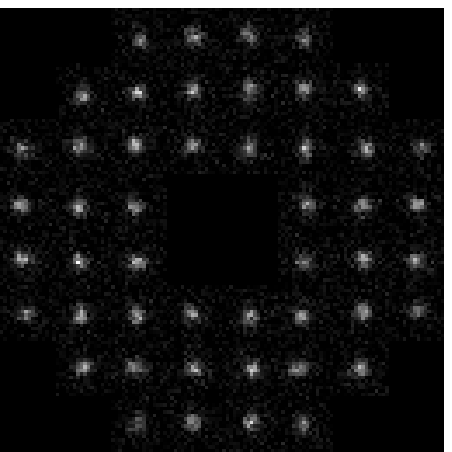}\hspace{0.1cm}%
\includegraphics[width=0.3\linewidth]{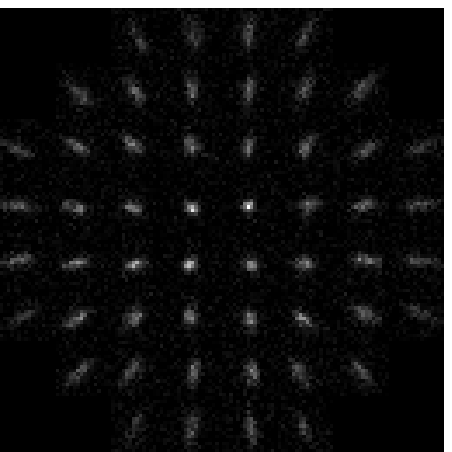}%
\includegraphics[width=0.3\linewidth]{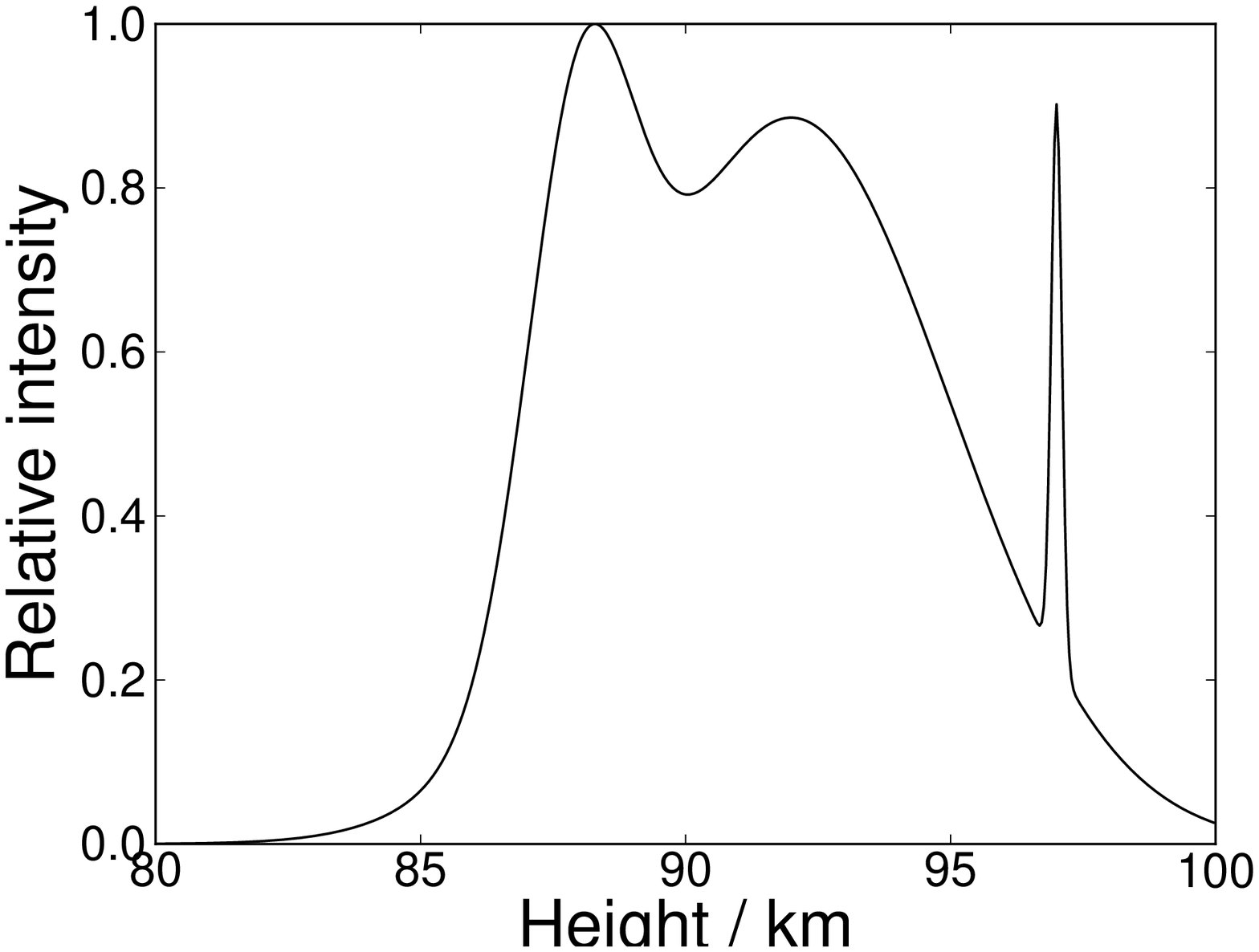}
\caption{(a) A figure showing a typical Shack-Hartmann spot pattern
  for a natural guide star. (b) For a laser guide star. (c) Showing a
  typical non-uniform sodium layer density profile \citep{2003A&A...403..775B}.}
\label{fig:shsspots}
\end{figure}

Most astronomical \ao systems use Shack-Hartmann wavefront sensors
\citep{shs}.  With \ngss, an array of Airy disc spots is produced
because the source is unresolved.  However with \lgss these spots are
elongated, as demonstrated in Fig.~\ref{fig:shsspots}, due to the
geometrical effect of viewing an extended source off-axis.

Shack-Hartmann wavefront sensors estimate the local wavefront gradient
or slope in each sub-aperture by measuring the offset of the
individual spots from a reference position defined by their position
when a flat wavefront is incident on the wavefront sensor.  A \cog
measurement can provide an essentially unbiased estimate (depending on
sampling, signal and background levels, and field of view) for
circularly symmetric \ngs spots.  However for elongated \lgs spots,
this is not the case if field of view or flux are limited (i.e.\ any
practical situation).  This is because signal and noise are spread
unevenly along the elongated and un-elongated directions, and hence
noise propagation is different along these directions which can affect
wavefront reconstruction.  Spot truncation in the direction of
elongation can also occur, 
resulting in biased slope measurements, particularly if the spot is under
sampled.  This is compounded for Sodium \lgss \citep{laserguidestar}
by the non-uniform profile of the Sodium layer density in the upper
atmosphere which causes Sodium \lgs spots to be non-smooth
(Fig.~\ref{fig:shsspots}), introducing complications for spot offset
measurements.  Likewise, for Rayleigh \lgss
\citep{1992OptL...17.1485T}, it can be necessary to increase the
optical range-gate depth to increase photon flux in poor seeing
conditions, leading to greater spot elongation, which can again lead
to complications for \wfs slope measurement.

A solution to the spot elongation problem is to compute the
correlation of the \lgs spots with a reference image
\citep{2006MNRAS.371..323T}, and then find the \cog of this
correlation image, thus giving the spot offset.  This naturally leads
to the question of how to obtain a suitable reference image, which
will be different for each sub-aperture.  This form of slope
estimation is commonly used for solar \ao \citep{2012SPIE.8447E..2NR},
where the Shack-Hartmann images have a very different structure, and
an identical reference for each sub-aperture can be used.  Correlation
is usually performed using a two dimensional Fast Fourier transform to
reduce computational complexity.  For night-time astronomical \ao,
the reference images differ in each sub-aperture due to the
geometrical nature of the spot elongation.

\subsection{Solar adaptive optics wavefront sensing}
Solar \ao systems use this correlation technique for wavefront
sensing.  The reference image used is identical for each sub-aperture,
and is typically chosen from a single sub-aperture in a single frame
of Shack-Hartmann images, updated approximately every 30 seconds.
This reference image updating is necessary because the Sun's surface
is constantly changing, and so the correlation between a given
Shack-Hartmann image and the reference decreases with time, reducing
the \ao system correction quality.  Updating the reference image in
this way will cause the science image to change position, since this
reference image changes for all sub-apertures, and thus introduces an
identical constant wavefront slope offset for each sub-aperture.
However this is not a problem for solar \ao because the integration
times for the science images are typically only a few seconds
(otherwise image saturation will occur).  Therefore, update of the
correlation reference image can be synchronised with a new science
image exposure.

\subsection{Key differences for astronomical correlation wavefront
  sensing} The situation for night-time astronomical \ao correlation wavefront
sensing is more complicated.  A single reference image cannot be used
because the shape of each Shack-Hartmann spot is different (defined in
part by the distance and angle of the sub-apertures from the laser
launch axis).  Therefore, a unique correlation image for each
sub-aperture must be supplied.

Astronomical \ao systems generally work in the photon-starved regime,
where fluxes and signal-to-noise ratios are low.  This is generally
also the case when using \lgss primarily because of the increasing
cost of more powerful lasers, the desire to operate with high frame
rates and at high \wfs order.  A single frame \lgs image is therefore
generally noisy, and so the computation of correlation
reference images cannot be obtained from a single
frame of Shack-Hartmann data since flux is low, and such images would
contain significant noise.  Rather, it is necessary to integrate
wavefront sensor images for long enough to provide a reference image
with a high signal to noise ratio, but over a time period short enough
to ensure that the Sodium layer density profile has not changed
significantly to avoid blurring of the reference images.  Wavefront
sensor images spanning of order one second of time would seem to be a
reasonable compromise (as we will show), though this will depend upon
the \ao system in question, the Sodium layer variability (which is not
constant) and the lasers.

Finally, science integration times for astronomical instruments are
far longer than those for solar observations.  Therefore it is not
possible to only update the correlation reference images in between
science exposures, since the time interval is too long and the
correlation reference images would quickly become out of date leading
to poor \ao performance.  

This paper presents a solution to these problems, with on-sky testing
reported using the CANARY demonstrator instrument on the William
Herschel Telescope. \ignore{canary}  This is of
crucial importance for future \elt \moao systems where wavefront
sensing is performed in open-loop.  When operating in open-loop,
wavefront sensors do not see the effects of changes made to \dm shape,
and therefore spot motions can be large.  Other wavefront sensing
algorithms, such as a weighted \cog or matched filter algorithms
\citep{2006MNRAS.371..323T} are non-linear and so work only poorly for
open-loop wavefront sensing.  The use of a correlation algorithm gives
significant performance advantages over a basic \cog algorithm
\citep{2006MNRAS.371..323T}, while still maintaining the linearity
essential for open-loop systems.

In \S2 we present a technique for updating the correlation reference
images in an astronomical \ao system whilst maintaining engagement of
the \ao loop, without affecting science image quality.  In \S3 we give
details of our implementation of this technique and results obtained
both from simulation and on-sky. \ignore{with CANARY.}  Finally in \S4 we draw
our conclusions.

\section{Real-time update of correlation reference images}
There are several techniques that can be used to obtain suitable
correlation reference images from noisy \wfs images, which we now discuss.

\subsection{Generation of correlation reference images}
\subsubsection{Integration of wavefront sensor images}
The first technique to obtain correlation reference images is to
simply sum a suitable number of \lgs wavefront sensor image frames,
for each sub-aperture.
The number of image frames should be chosen to achieve a good
signal-to-noise ratio, while remaining low enough that the Sodium
profile remains approximately constant during this time, and we
suggest that of order 0.1-1~s is appropriate.  However, the
drawback of this simple approach is due to the atmosphere itself,
which will cause blurring of the integrated image due to spot movement
between frames.  Therefore, the correlation reference images will not
be optimal.

\subsubsection{Shifting and adding}
The second technique that can be used to obtain correlation reference
images is to shift and add \wfs images from a number of individual
frames (again, probably spanning of order one second) on a
per-sub-aperture basis (i.e.\ with potentially a different shift for
each sub-aperture).  To use this technique, the position of each spot
in each frame is determined.  These spots are then shifted to the
position corresponding to zero (either using an integer pixel shift or
using sub-pixel interpolation), and summed with the corresponding
spots from other image frames.

The determination of the spot position can be made (on a
per-sub-aperture basis) either using a simple \cog
estimation, or by correlating with the current reference images, or by
correlating with a model of the spots.

\subsubsection{Modelling}
A further technique which can be used to obtain the reference images
is using modelling.  A small number of \lgs wavefront sensor images
can be obtained, and these used to derive parameters for a model of
the Sodium layer density profile, for example number and separation of sodium
density peaks.  Alternatively, an external profile monitor could be
used.  This model can then be used to generate suitable correlation
reference images.  We do not consider this modelling approach further
here, rather concentrating on data driven approaches.

\subsection{Updating the reference images}
\label{sect:update}
\begin{figure}
\includegraphics[width=\linewidth]{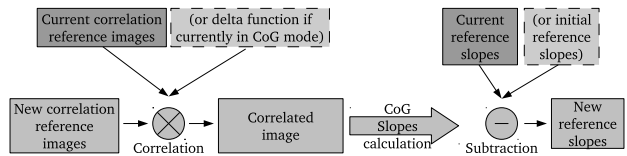}
\caption{A diagram showing the technique used to obtain reference
  slope measurements, and thus enable simultaneous updating of
  correlation reference images and reference slopes during closed-loop
  operation.}
\label{fig:corrRef}
\end{figure}

A flat incoming wavefront does not usually correspond to wavefront
slope measurements of zero.  Rather, it is necessary to perform a
calibration to measure a set of reference slope measurements that
correspond to a flat wavefront on the science image, leading to best
\ao corrected image improvement.  Changing the correlation reference
images will lead to different slope measurements for a given
wavefront, and therefore it is necessary to obtain a different set of
reference slopes for each set of correlation reference images.  If the
correlation reference images are uploaded to a real-time control
system without further considerations (i.e.\ without also modifying
the reference slopes), the \ao corrected science image \psf will
change each time, leading to poor performance.  This is because the
zero point (the estimated spot location for a flat input wavefront)
changes depending on the correlation reference.  The solution which we
propose is therefore to update the reference slope measurements at the
same time as the reference correlation images.  We here outline the
method by which these correlation reference slopes can be obtained,
which is also demonstrated in Fig.~\ref{fig:corrRef}.  This is key to
operating a correlation based wavefront sensor with a night-time
astronomical \ao system.

\subsection{Reference slope calculation}
We define the {\it ``Initial Reference Slopes''} as those that give the best
science \psf when a \cog centroiding algorithm is used.
Here, we note that correlation of an image with a two dimensional
delta function (i.e.\ an array of zeros, with a value of
one in the centre) yields the image itself, and so the \cog
centroiding algorithm yields the same result as correlation
with this identity followed by a \cog calculation.

We can compute the new reference slopes that will be required when the
first set of correlation reference images are to be used, as follows (assuming
that the system is previously using \cog or correlation with a delta
function).  We calculate the correlation of the new reference images
with a delta function.  The centroid measurements of the resulting
correlation images are then obtained, and these values subtracted from
the {\it ``Initial Reference Slopes''}, to give the new reference
slopes to be used alongside the new correlation reference images.
Strictly speaking the correlation is unnecessary here, however it has
been included for completeness, when comparing with the next section.

When a further update of correlation reference images is required, this
process can be repeated:  The correlation of the new images with a
delta function is calculated, and the centroid measurements of these
correlations obtained.  These centroid measurements are then
subtracted from the {\it ``Initial Reference Slopes''} (not the reference slopes
currently in use), giving the new reference slopes to be used with the
new correlation reference images.

\subsubsection{Initial Reference Slope ignorance}
\label{sect:noisepropagation}
It is entirely possible that the {\it ``Initial Reference Slopes''} are not
known.  This could be because the science \psf has been optimised with
correlation reference images already in use, because an automatic ongoing
optimisation of reference slopes is in place, or for other reasons.
In this case, it is not possible to follow the preceding procedure for
reference slope calculation, and some modifications are required as
follows.  

Rather than computing the correlation of the new reference images with
a delta function, the correlation of new and current reference images
is computed.  The centroid measurements of these correlations are
obtained, and these values subtracted from the reference slopes that
are currently in use (not the {\it ``Initial Reference Slopes''}), to
give the new reference slopes, which can then be used alongside the
new correlation reference images.

It should be noted that this scheme only relies on knowing the current
correlation reference images, and the current reference slope
measurements, both of which are easily obtainable from a real-time
control system.  No other state information is required (e.g.\ the
{\it ``Initial Reference Slopes''}), and so this technique is more robust
against operator error, for example the {\it ``Initial Reference Slopes''}
being out of date.

When using this scheme, it may be
possible for errors in reference slopes to build up over time,
something we investigate further in this paper (\S\ref{sect:buildup}).  This can arise
because each time a new set of reference slopes are computed, there is
a reliance on the new and current correlation reference images both of
which will have some noise, and so each successive set of reference
slopes will have increased noise.

\subsection{Background complications}
Accurate calculation of wavefront slopes from correlation images is
dependent on correct treatment of image background levels.  In solar
\ao, where the sub-aperture images are extended it is necessary to use
a windowing function, such as a Hamming window when performing the
correlation to reduce edge effects.  In astronomical \ao, such a
windowing function is not necessary if background levels are correctly
removed and the \shs spots do not fill the sub-apertures .  However, if
the background level is not correctly removed, or significant detector
noise is present, difficulties remain.  In this case, the correlation
image can have a significant background level, which will bias the
slope measurement unless treated correctly.

As a solution for this, we suggest the use of a brightest pixel
selection algorithm (\citet{basden10}(a)), for which the $n_\mathrm{i}$
brightest pixels in the $i^\mathrm{th}$ sub-aperture are selected for
further processing, and the background level is set at the value of
the $n_\mathrm{i}+1^\mathrm{th}$ brightest pixel in this sub-aperture.  This has
been shown to significantly reduce the effect of variable backgrounds
and the impact of readout noise.  The use of this algorithm is assumed
throughout this paper unless otherwise stated, and a value of 40 is
considered typical for $n_\mathrm{i}$ here, i.e.\ only the 40 brightest pixels in
each sub-aperture are selected for processing, the rest being set to
zero.  It should be noted that this is only sensible for sub-apertures
that contain more pixels than this.  By using this algorithm, we are
able to remove much of the background level from our correlation
images, allowing an improved estimation of wavefront slope to be made.

\subsection{Zero-padding and computational requirements}
\begin{figure}
\includegraphics[width=0.45\linewidth]{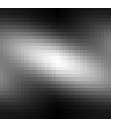}\hspace{0.1cm}%
\includegraphics[width=0.45\linewidth]{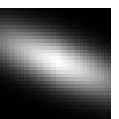}
\caption{(a) A correlation of a single sub-aperture with the
 corresponding correlation reference without zero padding.  The image
 wrapping can be clearly seen, affecting slope estimation. (b) As (a) but with the inclusion of zero padding.  No image wrapping occurs.}
\label{fig:zeropad}
\end{figure}

When Shack-Hartmann spots are extended with dimensions that are a
significant fraction of the sub-aperture size, it is necessary to
zero-pad the sub-aperture images before calculation of the image
correlation when using the Fast Fourier transform correlation method.
If this is not performed, the correlated image may wrap around at the
boundaries, producing a significant bias for slope estimation, as
shown in Fig.~\ref{fig:zeropad}. 

The computational complexity of a two dimensional Fast Fourier
transform scales as $\mathcal{O}(n^2\ln n)$ for images with side
dimension $n$.  Additionally, common Fast Fourier transform algorithms
perform better for certain image sizes, notably those
with a side length equal to a power of two.  Therefore, it is
advantageous for a real-time system to be able to have varying degrees
of zero-padding, depending on reference image size to reduce
computational requirements.  With a \lgs system, variation in
reference image spot size with distance from the laser launch axis
allows computational savings to be made by varying the sub-aperture
sizes, or amount of zero-padding, depending on expected elongation.
Once the correlation has been performed, the image size, and therefore
the computational effort in the \cog computation, can be reduced by
clipping the correlation images.

\subsection{Real-time control implementation}
Real-time control for \ao is computationally demanding, particularly
for \elt scale systems, since \wfs images must be processed and the
\dm demands computed within short timescales.  The use of correlation
wavefront sensing, requiring the computation of many small image
correlations in a short period of time further adds to the
computational demands placed on the real-time control system.  In this
section, we describe the real-time implementation of correlation
wavefront sensing that is implemented in the real-time control system
used with our on-sky demonstration \ignore{the CANARY} instrument.

\subsubsection{DARC}
\Darc \citep{basden9} is a high performance, \cpu based real-time
control system for \ao, which has been used successfully on-sky with
CANARY.  \darc has a modular design which affords it great flexibility
for developing and testing new algorithms.  An advanced correlation
wavefront sensing module has been developed, and used both on-sky and
for this investigation.  The suitability of \darc for use with
\elt-scale \ao systems has been demonstrated (\citet{basden11}(b)), and
includes the ability to make use of additional computational hardware
including \fpgas, \gpus and the Intel Xeon Phi, a many-core hardware
accelerator.

The \darc system has the ability to use rectangular sub-apertures that
are positioned and sized arbitrarily, i.e.\ a regular grid-like
pattern is not required.  This allows sub-apertures to be sized
according to expected spot elongation which varies across the
telescope pupil, thus reducing computational requirements.  The
correlation of each sub-aperture with an associated reference image is
performed using a Fast Fourier transform algorithm, with intermediate
steps stored in half-complex format (being Hermitian), thus reducing
memory bandwidth requirements.  Arbitrary zero-padding can be defined,
in either dimension, on a per-sub-aperture basis, allowing performance
savings to be made when spots are small relative to sub-aperture size,
and ensuring that aliasing of the Fast Fourier transforms does not
occur.

A (weighted) \cog algorithm is used to measure the centroid
of the correlation image, which can be clipped to reduce computational
requirements, particularly when zero-padding has been implemented.
Alternatively, a two dimensional parabolic or Gaussian fit to the
correlated image can also be used to determine wavefront slope, though
this is not discussed further here.

\subsubsection{Reference image and slope update}
\darc has the ability to allow synchronous update of parameters on a
per-frame basis.  Our graphical on-sky \ignore{The CANARY} correlation
tool allows reference images to be obtained (averaging calibrated \wfs
images), and from these, updates to the current reference slope
measurements are computed, as detailed in sections following
\S\ref{sect:update}.  The new reference images are then uploaded at
the same time as the new reference slopes, allowing the \ao loop to
remain closed (or engaged, for open-loop systems) while this operation
is performed.

\subsection{Sodium layer variability}
It is important to realise that this technique does not solve the
Sodium layer variability problem \citep{2006OptL...31.3369D}, which arises
because of uncertainty in the time-varying profile of Sodium layer
density.  However, what it does allow is for a continual optimisation
of \wfs performance by matching the reference images as closely as
possible to the true Sodium layer density profile.

In the case that an optical range gate can be applied to the Sodium
layer return signal (i.e.\ a pulsed laser and a fast \wfs shutter to only
image pulses of light returning from a certain height in the atmosphere), then
the uncertainty due to Sodium layer variability can be removed.  The
technique presented here then has a greater effect improving wavefront
slope measurements even in the presence of changing Sodium layer
density profiles.

\subsection{On-sky validation of reference update}
The technique proposed here for simultaneous update of correlation
reference images and slopes has been tested on-sky during the night
starting on 25th May 2013 using the \canary multiple \lgs \ao system.
\ignore{using CANARY.}  During these tests, with \canary operating in
a \moao mode, the reference images and slopes were updated while the
\ao loop remained active.  We recorded \ao corrected \psfs obtained
while using a \cog slope estimation algorithm, and then while using
the correlation algorithm, with four successive correlation reference
image updates.  Finally a further \ao corrected \psf was obtained
while using a \cog algorithm for slope estimation.

\begin{figure}
\includegraphics[width=\linewidth]{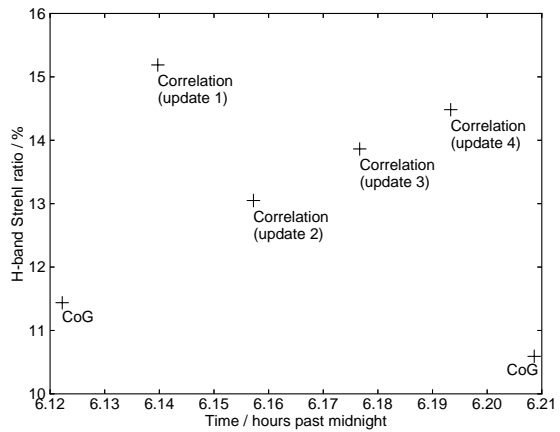}
\caption{A figure showing \ao corrected Strehl ratio as a function of
  time when using both a \cog algorithm, and correlation algorithm
  (with successive updates of the correlation reference images while
  the \ao loop was active).  These data were recorded using the
  \canary instrument on the night starting 25th May 2013 (so the data
  themselves are from the 26th May).}
\label{fig:canaryStrehl}
\end{figure}

Some variation in Strehl ratio is expected due to the variability of
atmospheric turbulence strength.  However, as shown in
Fig.~\ref{fig:canaryStrehl}, we do not see a continual worsening of
performance for consecutive reference updates.  We therefore conclude that our
technique of updating the correlation reference images on-sky works
well, thus mitigating some of the risk attached to \elt scale \lgs \ao
systems.  This figure spans a period of about five minutes close to
dawn, and the \ao performance obtained using correlation is shown to
be better than that using the \cog algorithm (taken both before and
after the four consecutive correlation updates).  The standard \canary
data processing tool was used to obtain these measurements.

\section{Reference image technique comparisons}
To investigate the correlation reference update techniques detailed in
the preceding sections, we have used the \dasp \citep{basden5}.  We
have interfaced  this simulation tool with \darc, providing a
hardware-in-the-loop simulation capability \citep{basden13}.  Our approach is as follows:
\begin{enumerate}
\item A sequence of noiseless, atmospherically propagated \lgs elongated \shs spot
  patterns is generated.
\item Photon shot noise, and detector readout noise is introduced to
  these images.
\item These images are passed into the real-time control system, \darc.
\item Correlation reference images are obtained using a number of
  these calibrated \wfs images, either the mean image, by shifting and
  adding, or by interpolating as described previously.
\item A new, independent, set of noiseless and noisy \lgs \shs spot
  patterns are generated.  The correlation algorithm is used to
  estimate the \wfs slope measurements using the noisy spot patterns,
  and these slope measurements are compared with the known (noiseless)
  slopes.  A \cog algorithm is also used with the noisy spot patterns
  for performance comparisons.
\end{enumerate}

\begin{figure}
% For some reason, libreoffice doesn't like converting this directly
% to eps - the shs image is missing.  So, save to svg, view in
% firefox, print to file, convert using ps2epsi, then change bounding
% box to:
% 19 540 554 752
\includegraphics[width=\linewidth,clip]{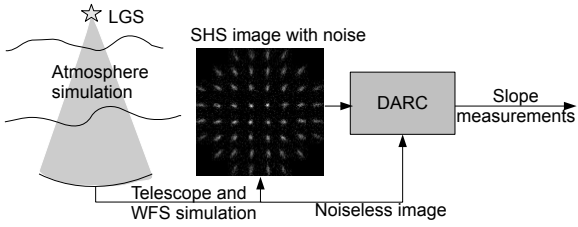}
\caption{A diagram showing our simulation approach.  Laser guide stars
  are propagated through the atmosphere, and Shack-Hartmann images
  formed, with photon noise and readout noise.  These images are
  passed to an instance of DARC, which processes them, providing
  wavefront slope measurements.  These slope measurements are then
  compared with the true slope measurements.}
\label{fig:overview}
\end{figure}

\begin{figure}
\includegraphics[width=0.24\linewidth]{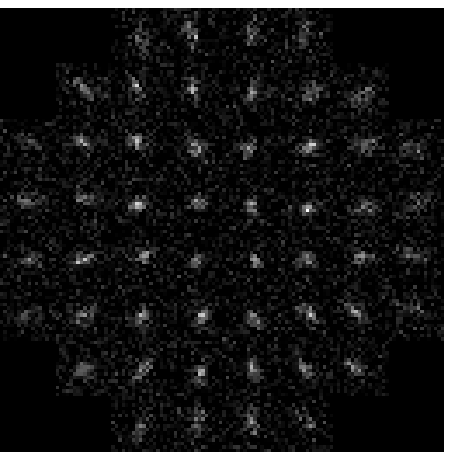}\hspace{0.1cm}%
\includegraphics[width=0.24\linewidth]{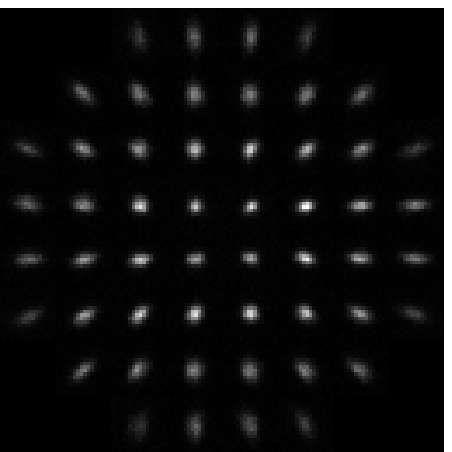}\hspace{0.1cm}%
\includegraphics[width=0.24\linewidth]{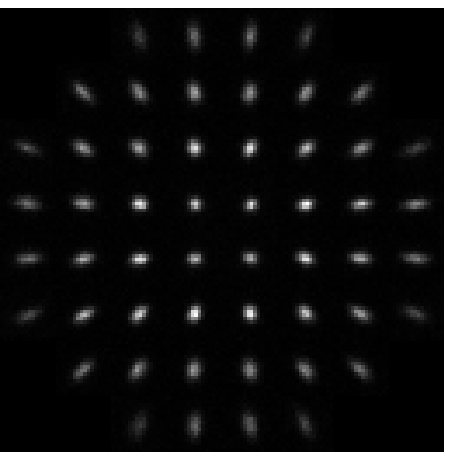}\hspace{0.1cm}%
\includegraphics[width=0.24\linewidth]{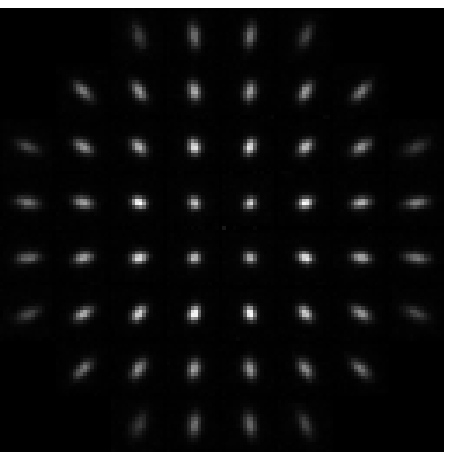}
\caption{(a) A figure showing a simulated Laser Guide Star Shack
  Hartmann image, with a mean signal of 200 photons per fully
  illuminated sub-aperture, and a read-out noise of two electrons.
  (b) A reference image computed from the sum of 100 SHS images.  (c)
  A reference image computed from 100 SHS images shifted and added.
  (d) A reference image computed from 100 SHS images interpolated and
  shifted, then summed.}
\label{fig:shs}
\end{figure}

Our performance metric is the \rms difference between estimated and
true slope measurements.  This technique is demonstrated in
Fig.~\ref{fig:overview}.  Although the visual difference between the
different techniques for reference image creation is small
(Fig.~\ref{fig:shs}), the following results show that the technique used
is worthy of consideration.

\subsection{Simulation parameters and method}
\dasp is a Monte-Carlo time-domain simulation code, which models the
atmosphere using a series of translating phase screens, and uses
physical optics for modelling the telescope and optical components.
Here, we simulate wavefront sensor images based on a $8\times8$
sub-aperture \shs on a 3.93~m telescope.  We use a 90~km laser beacon
with a Sodium layer depth of 30~km and a Gaussian profile.  Although
such a depth is unrealistic, it allows us to simulate the extreme
elongation that would be seen at the edge of an \elt instrument, as
well as less elongated spots from close to the laser launch position,
and therefore our results are relevant for these future telescopes.  A
simulated spot image is shown in Fig.~\ref{fig:shs}(a), covering a
range of spot elongations due to different off-axis viewing angle.
Unless otherwise stated, we do not vary the Sodium profile in our
simulations.  This does not significantly affect our results, because
we concentrate on short periods of time (typically less than a
second), over which the profile would not change significantly, and
also show that the correlation reference can be updated whilst the \ao
loop is closed without affecting performance, i.e.\ results will not
be affected when the profile does change.  Although \dasp is
suitable for \elt-scale simulation \citep{basden8,basden12}, it would
take a prohibitive amount of time to perform the investigations that
we carry out here at these scales.  We therefore approximate, using
sub-apertures of the same size as those expected on the \eelt (about
0.5~m), and spot elongations that cover the range of those expected on
the \eelt.

Unless otherwise stated, we assume a mean \wfs signal of 200 photons
per fully illuminated sub-aperture per frame, and a readout noise of
two electrons, and assume $16\times16$ pixels per sub-aperture.  We
model the atmosphere with a Von Karman model using an outer scale of
$L_0=50$~m, and a Fried's parameter of $r_0=10.6$~cm, and have a 4~ms
\wfs integration time.  We assume that 70\% of the turbulence is at the
ground layer, with 30\% at 11~km, though since we are not performing
any wavefront reconstruction, our results are not dependent on
atmospheric profile (though spot broadening will occur for worse seeing).  We use a wavefront sensor pixel scale of
0.25~arcsec per pixel, with a sub-aperture field of view of 4~arcsec.

Results are shown for the most elongated \lgs spots at $45^\circ$
unless otherwise stated, and we find that our results do not greatly
depend on degree of elongation.  For computation of the correlation
reference images, we use 100 \shs image frames, as found to
near-optimal in \S\ref{sect:hundredFrames} (unless stated
otherwise), and the slope estimation inaccuracies are computed using
an ensemble of 10000 frames.

\subsection{Investigation of results}

\subsubsection{Investigation of correlated image clipping}
\begin{figure}
\includegraphics[width=\linewidth]{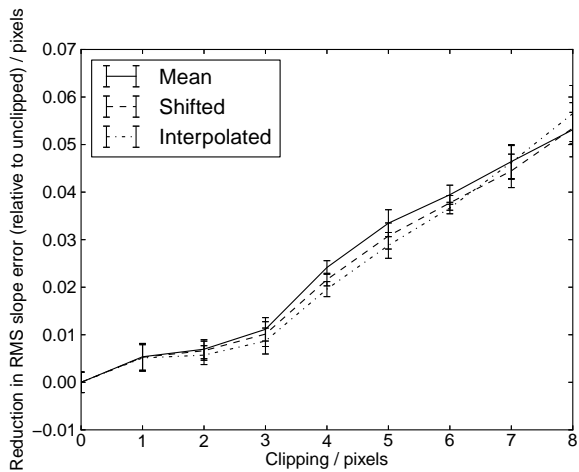}
\caption{A figure showing improvement in slope estimation as the
  clipping of correlated images is increased.  Here, the signal level
  is 200 photons per fully illuminated sub-aperture per integration,
  with a camera readout noise of 2 electrons.  The reduction in slope
  error is displayed relative to the error from unclipped correlated
  images, so that higher values signify improved slope estimation.}
\label{fig:clip}
\end{figure}

Oversampling of the reference images to avoid aliasing (wrapping)
leads to larger correlation images.  Clipping rows and columns from
the edges of these correlated images not only cuts out any noise
present in the edges, but also reduces computational requirements for
the resulting slope estimation.  Fig.~\ref{fig:clip} shows slope
estimation accuracy as a function of clipping (equal to the number of
rows and columns removed from each edge).  In this instance, a
clipping of eight returns the image to its pre-correlation size
($16\times16$ pixels), and is seen to give best slope estimation.  In
this case, the signal level is 200 photons per fully illuminated
sub-aperture per \wfs integration, and readout noise is two electrons.
At higher \snr levels, clipping does not lead to such a marked
improvement, however we recommend that clipping should be used as
performance is never seen to be worse.  100 \wfs images were used to
create the correlation reference images in this case.

\subsubsection{Investigation of background removal}
We have investigated background level subtraction using both a
conventional constant background level, and also using the brightest
pixel selection technique (with no prior background removal).

\begin{figure}
\includegraphics[width=\linewidth]{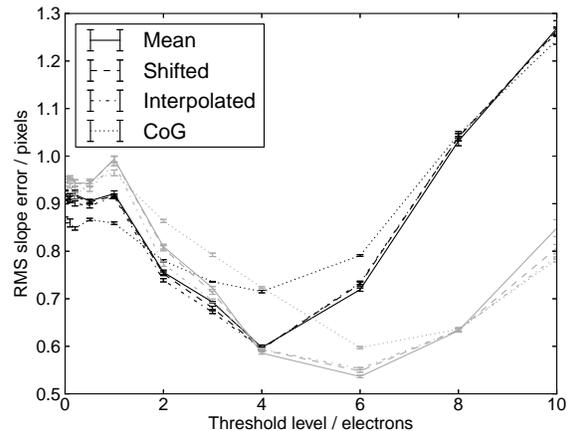}
\caption{A figure showing slope estimation error as a function of
  background level above mean background.  Here, the signal level is
  200 photons per fully illuminated sub-aperture per integration, with
  a camera readout noise of 2 electrons.  The black curves are for
  sub-apertures displaying large elongation, while grey curves are for
  sub-apertures close to the laser launch axis.  Lower values
  represent better performance.}
\label{fig:thresh}
\end{figure}

Fig.~\ref{fig:thresh} shows performance (slope error) of correlation
and \cog algorithms as a function of threshold level above the mean
background.  Here, the mean signal is 200 photons per pixel per
unvignetted sub-aperture, and the readout noise is two electrons.  The
threshold level is subtracted from all pixels, and any negative values
are then set to zero.  Best performance is seen with a threshold level
of two sigma (four electrons) above the mean, when using correlation
with elongated spots (far from the laser launch axis), while a higher
threshold level gives better performance for less elongated spots
(sub-apertures closer to the laser launch axis).  It is interesting to
note that if the threshold level is set too low for highly elongated
spots then the \cog algorithm performs better.

\begin{figure}
\includegraphics[width=\linewidth]{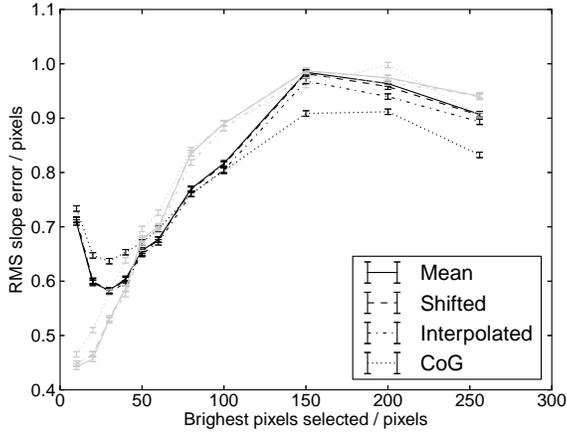}
\caption{A figure showing slope estimation error as the number of
  brightest pixels per sub-aperture selected for further processing is
  changed.  Here, the signal level is 200 photons per fully
  illuminated sub-aperture per integration, with a camera readout
  noise of 2 electrons.  The black curves represent highly elongated spots at the edge
  of the pupil, at 45$^\circ$, while the grey curves represent spots
  close to the launch axis, having less elongation.  Lower values
  represent better performance.}
\label{fig:ub}
\end{figure}

Fig.~\ref{fig:ub} shows slope estimation error as a function of number
of brightest pixels selected for further processing for a \wfs with a
mean of 200 photons per pixel per unvignetted sub-aperture, and a
detector readout noise of two electrons.  It has been shown that the
optimum number is dependent on spot size (\citet{basden10}(a)), and unless
otherwise stated through out this paper, we use a default value of 40,
which is close to minimum in this case, and provides near optimum
performance over the range of signal levels, readout noise and spot
sizes considered here.  Fig.~\ref{fig:ub} shows that correct
background subtraction leads to improved performance for correlation
techniques over \cog.  It should be noted that the background
subtraction based on brightest pixel selection provides better
performance than using a fixed background level, in part because it
provides an automatic way to vary background level across the \wfs
camera.

\subsubsection{Investigation of readout noise}
\begin{figure}
\includegraphics[width=\linewidth]{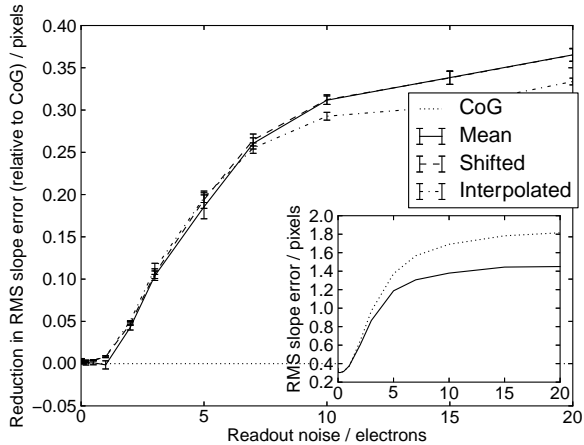}
\caption{A figure showing the reduction in slope estimation error
  relative to CoG as a function of detector readout noise, with higher
  values signifying improved slope estimation.  Inset is shown
  the actual slope error for the centre of gravity and correlation
  (using a mean reference) algorithm (here, higher values signify
  worse slope estimation).  The signal level is 200 photons per
  unvignetted sub-aperture per frame.}
\label{fig:noise}
\end{figure}

Increasing readout noise reduces \ao performance.  However, as shown
in Fig.~\ref{fig:noise}, correlation \wfs performance improves
relative to \cog performance as noise increases (although, of course,
slope estimation accuracy itself reduces).  This highlights the
benefits of correlation \wfs algorithms at lower signal-to-noise
ratios, and also shows that reference images produced using
interpolation provide better performance at the lowest signal-to-noise
ratios.
  
\subsubsection{Investigation of signal level}
\begin{figure}
\includegraphics[width=\linewidth]{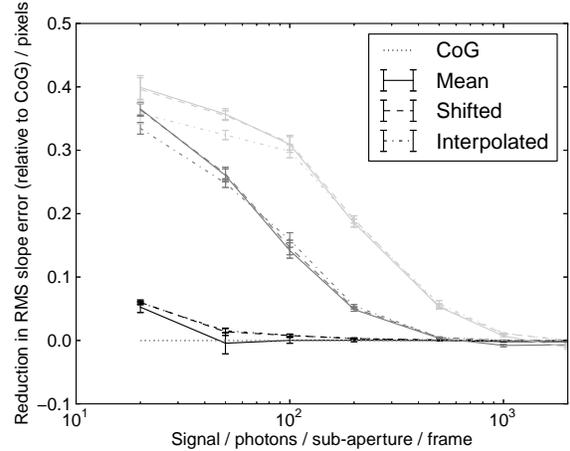}
\caption{A figure showing the reduction in slope estimation error
  relative to CoG as a function of WFS signal.  The readout noise is
  0.1 electrons (black curves), 2 electrons (dark grey curves) and 5
  electrons (light grey curves).  CoG slope error reduction is always
  zero since we are displaying reduction relative to CoG, and positive
values represent improved performance relative to CoG.}
\label{fig:sig}
\end{figure}

Fig.~\ref{fig:sig} shows the reduction in error of correlation \wfs
performance relative to \cog performance as a function of \wfs signal
level for different detector readout noise levels.  It can be seen
that at high light levels, the \cog and correlation performance is
similar, while at lower light levels, correlation provides better
performance than \cog.  It should be noted that shifted and interpolated
correlation reference images achieve better performance than
correlation reference images obtained from the mean of the pixel data.

\subsubsection{Investigation of number of reference frames}
\label{sect:hundredFrames}
\begin{figure}
\includegraphics[width=\linewidth]{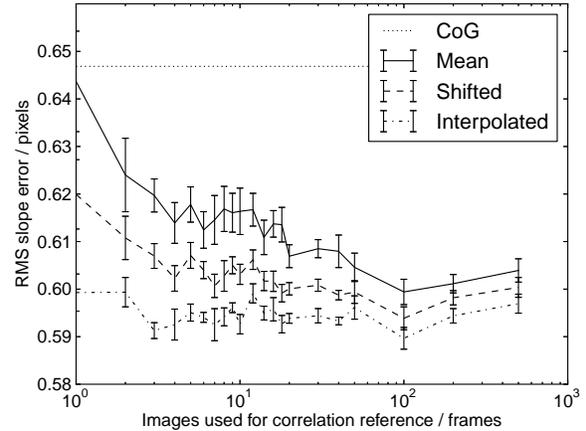}
\caption{A figure showing slope estimation error as a function of
  number of images used to build the reference image.  Here, the mean
  signal is 200 photons per sub-aperture with 2 electrons readout
  noise.}
\label{fig:source}
\end{figure}

The number of \shs image frames used to compute the correlation
reference images is of interest, regardless of the algorithm used to
compute the reference from the raw images.  Fig.~\ref{fig:source}
shows slope estimation error as a function of number of image frames
used to compute the reference image.  Here it can be seen that a
sample of approximately 100 images gives best performance, though
there is a fair degree of tolerance, particularly when using the
interpolated reference calculation technique.  The line for \cog is
also shown for comparative purposes (since this does not require
reference images).  The \rms slope error is seen to increase after
about 100 frames and this is due to increased spot blurring in the
reference images, resulting in reference spots that resemble the
actual median sub-aperture image less well.

\subsubsection{Investigation of threshold on correlated image}
\begin{figure}
\includegraphics[width=\linewidth]{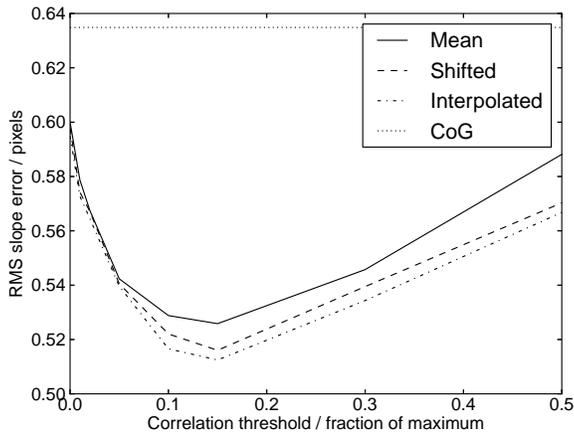}
\caption{A figure showing slope estimation error as a function of
  correlation image threshold level.}
\label{fig:corrthresh}
\end{figure}

After the correlation of a \shs spot with the reference image has been
computed, the resulting correlation image requires processing to yield
the local wavefront slope estimate.  Here we investigate the effect
that applying a threshold to this correlated image has on the slope estimation
accuracy.  Fig.~\ref{fig:corrthresh} shows slope estimation error as a
function of the correlation threshold, where the threshold is given as
a fraction of the maximum value in the correlation image.  Here, we
see that by selecting a threshold level that is a small fraction
(0.1--0.2) of the maximum value in each sub-aperture we can improve
the slope estimation accuracy.  

It should be noted that this is a separate threshold level than that
applied during image calibration.  Applying a threshold to the
correlated image helps to remove the effect of unwanted frequency
components, increasing slope estimation accuracy.  This also helps to
negate the effect of an incorrect image calibration (i.e.\ an
incorrect threshold applied to the raw image. 

\subsection{Reference slope noise propagation}
\label{sect:buildup}
\begin{figure}
\includegraphics[width=\linewidth]{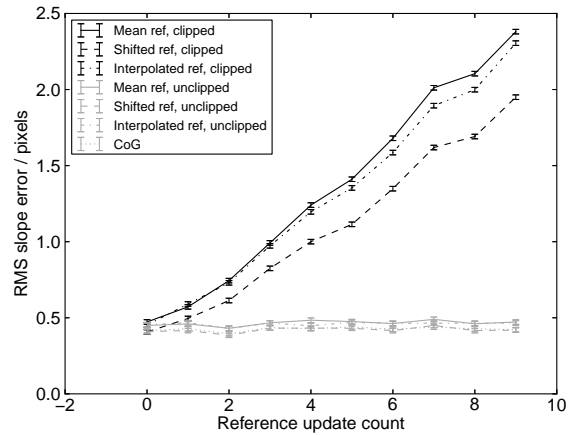}
\caption{A figure showing the error in slope
  measurements due to build up of bias in reference slopes when
  incorrect clipping is used.  Here, the x-axis represents the number
  of times the reference image and slopes have been updated.  The
  ``clipped'' results (black curves) show the bias introduced by
  incorrect clipping, while the grey curves show that it is possible
  to continue updating the references without affecting performance.}
\label{fig:clippropagation}
\end{figure}

As discussed in \S\ref{sect:noisepropagation}, it is possible (and
sometimes necessary) to update correlation reference images and
reference slopes using only a knowledge of the current system state.
One major source of bias in reference slope estimation comes from
clipping of the correlation reference images.  It is essential that
reference slope offsets are computed using the unclipped correlation
of the current and new reference images.  If this is not the case,
i.e.\ the slope offsets are computed using a clipped correlation image,
errors in these slope estimates quickly build up, as shown in
Fig.~\ref{fig:clippropagation}.  When computed correctly, we do not
see errors build up, even over long timescales.  

\section{Conclusions}
We have presented a technique that allows correlation reference images
to be updated in real-time for astronomical \ao systems, whilst the
\ao loop is closed, without affecting the science \psf.  This is of
particular interest for systems with \lgss, where spot patterns show
large amounts of elongation along one axis.  This technique has been
demonstrated successfully on-sky using the CANARY \ao
demonstrator instrument.  We have investigated techniques to improve
the accuracy of wavefront slope estimation related to the computation
of the image correlations, using a Monte-Carlo \ao simulation tool
interfaced to a real-time control system.  We recommend that of order
100 calibrated image frames are used for computation of the
correlation reference, and that these images should be calibrated
using a brightest pixel selection technique.  Our results show that
there is little difference between the three techniques investigated
for reference image computation and therefore recommend that the
shift-and-add method should be used since it performs well in most
situations.  We have also shown that these correlation based
algorithms are more accurate than a centre-of-gravity algorithm in low
and moderate signal-to-noise ratio regimes.  We have not investigated
truncation effects, always assuming that our elongated spots fit
within the sub-aperture field of view, though we will study this in a
future paper.

\section*{Acknowledgements}
This work is funded by the UK Science and Technology Facilities
Council, grant ST/I002871/1.  The author would like to thank the
CANARY team:  The original version of this paper did not contain
CANARY results, however the referee requested it be included.

\bibliographystyle{mn2e}

\bibliography{mybib}

\begin{thebibliography}{}

\bibitem[\protect\citeauthoryear{{Babcock}}{{Babcock}}{1953}]{adaptiveoptics}
{Babcock} H.~W.,  1953, \pasp, 65, 229

\bibitem[\protect\citeauthoryear{{Basden}, {Geng}, {Myers} \&
  {Younger}}{{Basden} et~al.}{2010}]{basden9}
{Basden} A.,  {Geng} D.,  {Myers} R.,    {Younger} E.,  2010, Appl.\ Optics,
  49, 6354

\bibitem[\protect\citeauthoryear{{Basden}, {Myers} \& {Butterley}}{{Basden}
  et~al.}{2010}]{basden8}
{Basden} A.,  {Myers} R.,    {Butterley} T.,  2010, Appl.\ Optics, 49, G1

\bibitem[\protect\citeauthoryear{Basden}{Basden}{2014}]{basden13}
Basden A.~G.,  2014, In press

\bibitem[\protect\citeauthoryear{Basden, Bharmal, Myers, Morris \&
  Morris}{Basden et~al.}{2013}]{basden12}
Basden A.~G.,  Bharmal N.~A.,  Myers R.~M.,  Morris S.~L.,    Morris T.~J.,
  2013, \mnras, 435, 992

\bibitem[\protect\citeauthoryear{{Basden}, {Butterley}, {Myers} \&
  {Wilson}}{{Basden} et~al.}{2007}]{basden5}
{Basden} A.~G.,  {Butterley} T.,  {Myers} R.~M.,    {Wilson} R.~W.,  2007,
  Appl.\ Optics, 46, 1089

\bibitem[\protect\citeauthoryear{{Basden} \& {Myers}}{{Basden} \&
  {Myers}}{2012}]{basden11}
{Basden} A.~G.,  {Myers} R.~M.,  2012, \mnras, 424, 1483

\bibitem[\protect\citeauthoryear{{Basden}, {Myers} \& {Gendron}}{{Basden}
  et~al.}{2012}]{basden10}
{Basden} A.~G.,  {Myers} R.~M.,    {Gendron} E.,  2012, \mnras, 419, 1628

\bibitem[\protect\citeauthoryear{{Butler}, {Davies}, {Redfern}, {Ageorges} \&
  {Fews}}{{Butler} et~al.}{2003}]{2003A&A...403..775B}
{Butler} D.~J.,  {Davies} R.~I.,  {Redfern} R.~M.,  {Ageorges} N.,    {Fews}
  H.,  2003, \aap, 403, 775

\bibitem[\protect\citeauthoryear{{Davis}, {Hickson}, {Herriot} \&
  {She}}{{Davis} et~al.}{2006}]{2006OptL...31.3369D}
{Davis} D.~S.,  {Hickson} P.,  {Herriot} G.,    {She} C.-Y.,  2006, Optics
  Letters, 31, 3369

\bibitem[\protect\citeauthoryear{{Foy} \& {Labeyrie}}{{Foy} \&
  {Labeyrie}}{1985}]{laserguidestar}
{Foy} R.,  {Labeyrie} A.,  1985, \aap, 152, L29

\bibitem[\protect\citeauthoryear{{Gendron}, {Vidal}, {Brangier}, {Morris},
  {Hubert}, {Basden}, {Rousset} \& {Myers}}{{Gendron}
  et~al.}{2011}]{canaryresultsshort}
{Gendron} E.,  {Vidal} F.,  {Brangier} M.,  {Morris} T.,  {Hubert} Z.,
  {Basden} A.,  {Rousset} G.,    {Myers} R.,  2011, \aap, 529, L2

\bibitem[\protect\citeauthoryear{{Richards}}{{Richards}}{2012}]{2012SPIE.8447E..2NR}
{Richards} K.,  2012, in Society of Photo-Optical Instrumentation Engineers
  (SPIE) Conference Series Vol.~8447 of Society of Photo-Optical
  Instrumentation Engineers (SPIE) Conference Series, {Adaptive optics real
  time processing design for the advanced technology solar telescope}

\bibitem[\protect\citeauthoryear{{Shack}}{{Shack}}{1971}]{shs}
{Shack} R.~V.,  1971, Journal of the Optical Society of America, 61, 656

\bibitem[\protect\citeauthoryear{{Thomas}, {Fusco}, {Tokovinin}, {Nicolle},
  {Michau} \& {Rousset}}{{Thomas} et~al.}{2006}]{2006MNRAS.371..323T}
{Thomas} S.,  {Fusco} T.,  {Tokovinin} A.,  {Nicolle} M.,  {Michau} V.,
  {Rousset} G.,  2006, \mnras, 371, 323

\bibitem[\protect\citeauthoryear{{Thompson} \& {Castle}}{{Thompson} \&
  {Castle}}{1992}]{1992OptL...17.1485T}
{Thompson} L.~A.,  {Castle} R.~M.,  1992, Optics Letters, 17, 1485

\end{thebibliography}
\bsp

\end{document}